\documentclass{article}
\usepackage[colorlinks, pdfborder={0 0 0}, plainpages=false]{hyperref}
\usepackage{graphicx}
\usepackage{grffile}
\usepackage{subfigure}
\usepackage{xspace}
\usepackage[usenames,dvipsnames]{color}
\usepackage{amssymb}
\usepackage[normalem]{ulem} 
\usepackage{letltxmacro}
\usepackage{amsfonts}
\usepackage{amsmath}
\usepackage{amssymb}
\usepackage{bm}
\usepackage{dcolumn}
\usepackage{epsfig}
\usepackage{graphicx}
\usepackage{graphics}
\usepackage[latin1]{inputenc}
\usepackage{latexsym}
\usepackage{rotating}
\usepackage{hyperref}
\usepackage[usenames]{color}
\usepackage{float}
\usepackage{xspace} 
\usepackage{mathrsfs}
\usepackage{subfigure}
\usepackage{multirow}
\usepackage{mathrsfs}
\usepackage{latexsym}
\usepackage{subfigure, graphicx}
\usepackage{caption}

\usepackage[T1]{fontenc}    
\usepackage{hyperref}       
\usepackage{url}            
\usepackage{booktabs}       
\usepackage{amsfonts}       
\usepackage{nicefrac}       
\usepackage{microtype}      
\usepackage[nonatbib,final]{nips_dlps_2017}

\newcommand*\samethanks[1][\value{footnote}]{\color{red} \footnotemark[#1]}

\title{Glitch Classification and Clustering for LIGO \\with Deep Transfer Learning}

\author{
	Daniel George~\thanks{Equal contribution} \\
	NCSA and Department of Astronomy \\
	University of Illinois at Urbana-Champaign \\
	 \texttt{dgeorge5@illinois.edu} \\
	 \And
	Hongyu Shen~\samethanks \\
	NCSA and Department of Statistics \\
	University of Illinois at Urbana-Champaign \\
	\texttt{hongyu2@illinois.edu} \\
    \And
	 E.~A. Huerta \\
		NCSA \\
	University of Illinois at Urbana-Champaign \\
	\texttt{elihu@illinois.edu} 
}

\begin{document}

\maketitle

\begin{abstract}
The detection of gravitational waves with LIGO and Virgo requires a detailed understanding of the response of these instruments in the presence of environmental and instrumental noise. Of particular interest is the study of anomalous non-Gaussian noise transients known as glitches, since their high occurrence rate in LIGO/Virgo data can obscure or even mimic true gravitational wave signals. Therefore, successfully identifying and excising glitches is of utmost importance to detect and characterize gravitational waves. In this article, we present the first application of \textit{Deep Learning} combined with \textit{Transfer Learning} for glitch classification, using real data from LIGO's first discovery campaign labeled by Gravity Spy, showing that knowledge from pre-trained models for real-world object recognition can be transferred for classifying spectrograms of glitches. We demonstrate that this method enables the optimal use of very deep convolutional neural networks for glitch classification given small unbalanced training datasets, significantly reduces the training time, and achieves state-of-the-art accuracy above 98.8\%. Once trained via transfer learning, we show that the networks can be truncated and used as feature extractors for unsupervised clustering to automatically group together new classes of glitches and anomalies. This novel capability is of critical importance to identify and remove new types of glitches which will occur as the LIGO/Virgo detectors gradually attain design sensitivity.

\end{abstract}

\section{Introduction}
\label{intro}

The LIGO~\cite{LSC:2015} and Virgo~\cite{Virgo:2015} detectors are the largest and most sensitive interferometric detectors ever built. They can even sense changes thousands of times smaller than the diameter of a proton~\cite{LSC:2015,DII:2016}. These instruments have already detected multiple gravitational wave (GW) signals produced from coalescence of black holes~\cite{GW1,GW2,thirddetection,GW4} as well as neutron star mergers~\cite{GW5,GWMMA}. As LIGO/Virgo gradually attain design sensitivity, they will transition into an astronomical observatory that will routinely detect new GW sources, providing insights into astrophysical events that cannot be seen through any other means~\cite{LVCT,SathyaLRR:2009}. 

For LIGO/Virgo to realize their full potential, it is necessary to ensure that their sensing capabilities are not hindered by unwanted non-Gaussian noise transients, known as glitches, which contaminate GW data. There are extensive ongoing efforts on separating glitches from signals, and/or classifying them based on their characteristics, which is a non-trivial task requiring ``intelligent'' algorithms given that the glitches vary widely in duration, frequency range and morphology, spanning a wide distribution that is challenging to model accurately~\cite{corn:2015CQGra,jade:2015CQGra,jade1:2016,DBNN,DNN,DNN2,DeepTransfer}. Furthermore, since the LIGO/Virgo detectors are undergoing commissioning between each observing run, we expect that new types of glitches will be identified as they attain reach sensitivity~\cite{DII:2016,D7:2016,D8:2016}. 

Accurately classifying glitches is essential for several reasons~\cite{jade1:2016,spy:2016arXiv}: a) This will prevent false GW detections due to coincident glitches across multiple detectors that closely mimic signals. b) Rapidly identifying and excising glitches will enhance detector sensitivity, and improve the significance of GW signals that are contaminated by glitches~\cite{GW5}. c) The LIGO/Virgo detectors have thousands of instrumental and environmental channels to monitor changes that occur due to environmental or hardware issues. By carefully tracking down these glitches, we aim to identify their source and eliminate them promptly to ensure that the data stream is usable for GW data analysis.

The complex and time-evolving nature of glitches makes them an ideal case study to apply machine learning algorithms. Deep learning algorithms have been recently applied for GW signal detection and parameter estimation~\cite{DNN2NIPS} as well as for denoising LIGO data~\cite{Denoising}. In this article, we focus on deep learning with Convolutional Neural Networks (CNNs)~\cite{DL-Nature} for glitch classification, using spectrogram images computed from the time-series data as inputs. Recent efforts on this front include \texttt{Gravity Spy}, an innovative interdisciplinary project that provides an infrastructure for citizen scientists to label datasets of glitches from LIGO via crowd-sourcing~\cite{spy:2016arXiv}. Supervised classification algorithms based on this dataset were presented in the first Gravity Spy article~\cite{spy:2016arXiv} and were further discussed in~\cite{multi:2017arXiv}. These algorithms employed deep learning CNN models which were 4 layers deep, and achieved overall accuracies close to 97\% for glitch classification. It was found, however, that glitch classes with very few labeled samples could not be classified with the same level of accuracy.

Here, we present \textit{Deep Transfer Learning} as a new method for glitch classification that leverages pre-trained state-of-the-art CNNs used for object recognition, and fine-tunes them throughout all layers to accurately classify glitches after re-training on a small dataset of LIGO spectrograms. We show that this technique achieves state-of-the-art results for glitch classification with the \texttt{Gravity Spy} dataset, attaining above 98.8\% overall accuracy and perfect precision-recall on 8 out of 22 classes, while significantly reducing the training time to a few minutes. Our results indicate that new types of glitches can be classified accurately given very few labeled examples with this technique. We also demonstrate that features learned from real-world images by very deep CNNs are directly transferable for the classification of spectrograms of time-series data from GW detectors, and possibly also for spectrograms in general, even though the two datasets are very dissimilar. The CNNs we use were originally designed for over 1000 classes of objects in ImageNet. Therefore, our algorithm can be easily extended to classify hundreds of new classes of glitches in the future, especially since this transfer learning approach requires only a few labeled examples of a new class. Furthermore, we outline how new classes of glitches can be automatically grouped together by using our trained CNNs as feature extractors for unsupervised or semi-supervised clustering algorithms.

\begin{figure*}
	\centering     
	{\includegraphics[width=.999\textwidth]{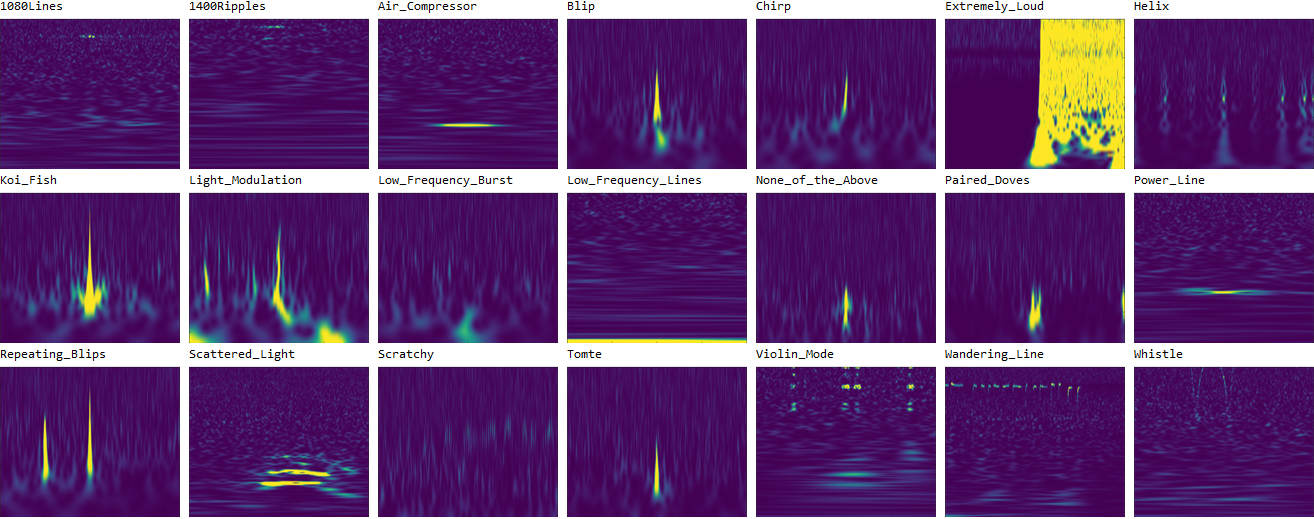}\vspace{.1in}}		
	\caption{Classes of glitches in the \texttt{Gravity Spy} dataset from the first observing run of LIGO. The \texttt{No\_Glitch} class (periods of usual LIGO noise without transient glitches) is omitted in the figure. True GW signals from compact binary coalescence fall in the \texttt{Chirp} class. For each input, spectrograms with durations of 0.5s, 1.0s, 2.0s, and 4.0s are available. The objective is to predict the classes given the images.}
	\label{fig:glitches}
\end{figure*}

\section{Methods}
\label{met}


The \texttt{Gravity Spy} crowd-sourcing project mobilizes citizen scientists to hand-label spectrograms obtained from LIGO time-series data after being shown only a few examples, which indicates that generic pattern recognizers developed in humans for real-world object recognition are also useful when distinguishing spectrograms of glitches. This motivated us to apply a similar approach referred as ``transfer learning'' in the machine learning literature.

Transfer learning is an essential ingredient for true artificial intelligence, where knowledge learned in one domain for some task (typically where there is a large amount of labeled data) can be transferred to another domain for a different task~\cite{Transferable} where there may only be limited number of labeled examples. In the context of deep learning for classifying images, transfer learning can be performed by pre-training a deep CNN on a large and diverse dataset with well-established labels followed by modifying the final layer based on the number of required classes and then fine-tuning the weights on a different dataset of interest. It is well known that the initial layers of a CNN always learn to extract simple generic features (e.g., edges, corners, etc.), which are applicable to all types of images, whereas the final layers represent highly abstract and data-specific features~\cite{CNNFeatures}. Therefore, transfer learning is expected to result in higher accuracy and a faster training process compared to training the same CNNs from scratch, due to the shared features present in the initial layers.

To demonstrate the power of transfer learning for classifying glitches, we compare the performance of the most popular CNN models for object recognition, namely Inception~\cite{arxiv:Sz} version 2 and 3, ResNet~\cite{arxiv:Km}, and VGG~\cite{arxiv:karen}, all of which were leading entries in recent ILSVRC competitions. These CNNs were pre-trained on a large dataset of images --- i.e., \texttt{ImageNet}~\cite{cvpr:Deng}, which contains 1.2 million labeled images of real-world objects belonging to 1000 categories --- over the course of 2 to 3 weeks using multiple GPUs by other research groups. We obtained the open-source weights from these models, and used them to initialize the CNNs, before fine-tuning (re-training) each model on our training dataset of glitches. 

The \texttt{Gravity Spy} dataset, from the first observing run of LIGO, contains labeled spectrogram samples from 22 classes of glitches shown in Figure~\ref{fig:glitches}. We randomly split this dataset, containing about 8500 elements, into two parts such that approximately 80\% of samples in each class was in the training set, and 20\% of each class was in the testing set. These images were hand-labeled by citizen scientists participating in the \texttt{Gravity Spy} project, and the accuracy of the labeling was greatly enhanced by cross-validation techniques within the \texttt{Gravity Spy} infrastructure, also involving experts from the LIGO Detector Characterization team~\cite{spy:2016arXiv}.

The final fully-connected layer in each CNN model was replaced with another fully-connected layer having 22 neurons corresponding to each glitch class. The softmax function is used as the final layer in each model to provide probabilities of each class as the outputs. We fine-tuned across all the layers since the dataset of glitches is very different from the objects in the ImageNet data.

\section{Results}
\label{exp}

Both InceptionV2 and InceptionV3 achieved over 98\% accuracy in fewer than 10 epochs of training (less than 20 minutes), VGG16 and VGG19 achieved over 98\% accuracy within 30 epochs of training. In Table~\ref{results}, we compare the results of these CNNs trained with the transfer learning method and that of the CNN models described in~\cite{spy:2016arXiv,multi:2017arXiv} which were trained from scratch on the same training set for sufficient number of epochs.  We found that our models consistently achieved over 98\% accuracy for many epochs, thus indicating that the performance is robust, regardless of the stopping criteria, and therefore the model is not overfitting on the test set. Note that our models consistently under-performed with less than 98\% accuracy when trained without transfer learning. 

With InceptionV3, we achieved \textit{perfect} precision and recall on 8 classes: \texttt{1080Lines}, \texttt{1400Ripples}, \texttt{Air\_Compressor}, \texttt{Chirp}, \texttt{Helix}, \texttt{Paired\_Doves}, \texttt{Power\_Line}, and \texttt{Scratchy}.  With ResNet50, we achieved perfect precision and recall on 7 classes: \texttt{1080Lines}, \texttt{1400Ripples}, \texttt{Extremely\_Loud}, \texttt{Helix}, \texttt{Paired\_Doves}, \texttt{Scratchy}, and \texttt{Violin\_Mode}. Both ResNet50 and InceptionV3 achieved the highest accuracy of 98.84\% on the test set despite being trained independently via different methods on different splits of the data. Both models obtained 100.00\% accuracy when considering the top-5 predictions, which implies that given any input, the true class can be narrowed down to within 5 classes with 100.00\% confidence. This is particularly useful, since the true class of a glitch is often ambiguous to even human experts.

\begin{table}[t]
	
	\caption{\label{results} Accuracy of Classification on the Test Set}
	\footnotesize
	\begin{center}
		\begin{tabular}{llllll} 
		\toprule
			Neural Network  & Top-1 & Top-2 & Top-3 & Top-4 & Top-5 \\  
		\midrule
			
			Tuned-InceptionV3& 98.84\% & 99.71\% & 99.88\% & 99.94\% & 100.00\%  \\
			Tuned-InceptionV2& 98.78\% & 99.59\% & 99.71\% &  99.94\% & 100.00\%  \\
			Tuned-ResNet50& 98.84\% & 99.71\% & 99.83\% & 99.94\% & 100.00\%  \\
			Tuned-VGG16& 98.15\% & 99.36\% & 99.71\% & 99.83\% & 99.88\% \\
			Tuned-VGG19& 98.21\% & 99.31\% & 99.60\% & 99.71\% & 99.71\%  \\
			CNN in~\cite{spy:2016arXiv,multi:2017arXiv} & 96.70\%&98.32\%&99.13\%&99.31\%&99.36\% \\
		\bottomrule
		\end{tabular}
	\end{center}
\vspace{.1in}
	 This table lists the top-1 to top-5 accuracies for different CNNs on the testing set. We re-trained the 4 layer merged-view CNN model described in the publications~\cite{spy:2016arXiv,multi:2017arXiv} from scratch on our same train-test dataset for a fair comparison. Each model was trained with sufficient numbers of iterations until the error on the validation set started increasing. Note that our Inception and ResNet models are capable of narrowing down any input to within 5 classes with 100.00\% accuracy.

\end{table}
\normalsize 

\begin{figure*}
	\centering     
	{\includegraphics[width=.461\textwidth]{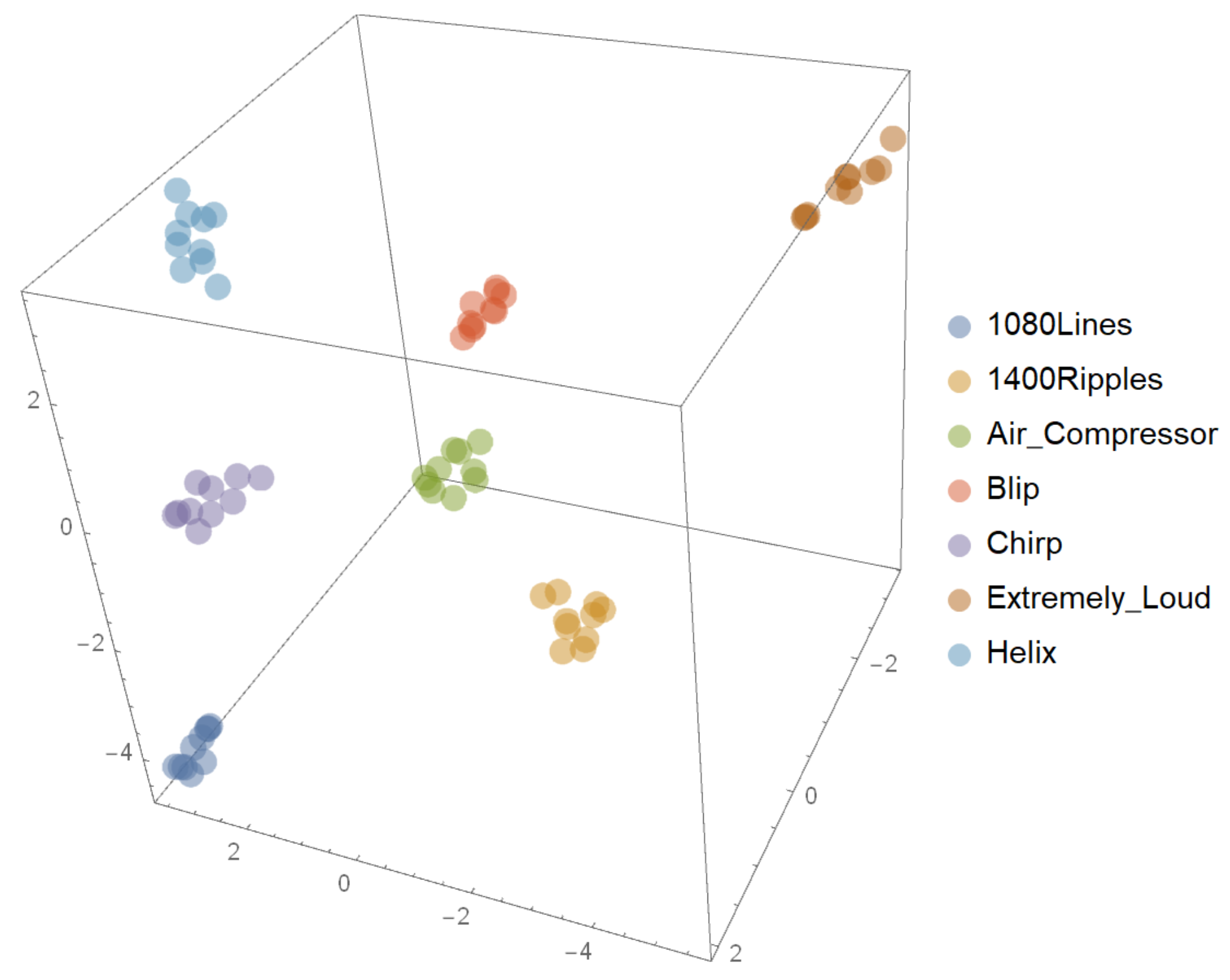}\hspace{.1in}
	\includegraphics[width=.515\textwidth]{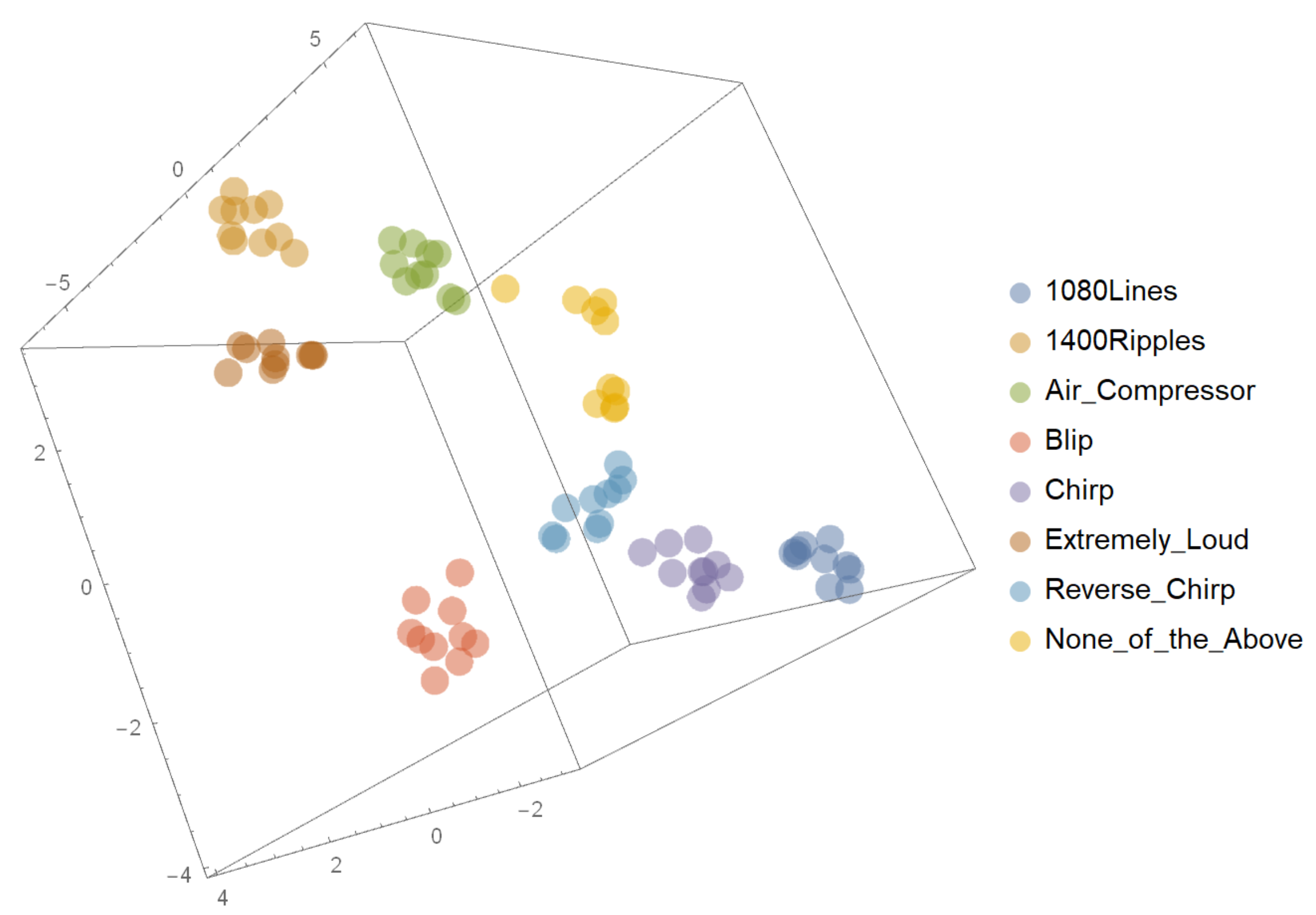}\vspace{.1in}}		
	\caption{\footnotesize Location of different classes of glitches from the test set after applying the CNN feature-extractor from our fine-tuned InceptionV3 model. The t-SNE~\cite{TSNE} algorithm was used to reduce the dimension to a 3D vector. Note that each type of glitch forms a cluster, and their relative positions depend on their morphology. Outliers may be inspected closely to verify their labels and decide whether they should belong to a new class. A new class called \texttt{Reverse\_Chirp} was added. It can been seen that the CNN feature-extractor maps this class (which was not shown during training) to a unique cluster. Furthermore, this cluster is located near the \texttt{Chirp} class and the \texttt{None\_of\_the\_Above} class, which means that the relative positions of the glitches in this feature-space is meaningful. Note that glitches in the \texttt{None\_of\_the\_Above} are also grouped into smaller clusters.}
	\label{fig:tSNE1}
\end{figure*}

We found that the trained CNNs may also be used as good feature extractors for finding new categories of glitches from unlabeled data in an unsupervised or semi-supervised manner. This method can be used to identify many more categories of noise transients and estimate at what times new types of glitches with similar morphologies start occurring. This may also be used to correct mislabeled glitches in the original dataset used for training/testing by searching for anomalies in the feature-space. For any of our models, removing the final softmax and fully-connected layer near the output produces a CNN that maps any input image to a vector of real numbers which encode useful information distinguishing different classes of glitches. In this high-dimensional space, glitches having similar morphology will be clustered together.  Therefore, when new types of glitches appear, which are classified as \texttt{None\_of\_the\_Above} by our CNN model, they may be mapped to vectors using these truncated CNN feature extractors (see Figure~\ref{fig:tSNE1}) and new clusters (classes) can be found. 

\begin{figure*}
	\centering     
	\hspace{-.04in}\includegraphics[width=.969\textwidth]{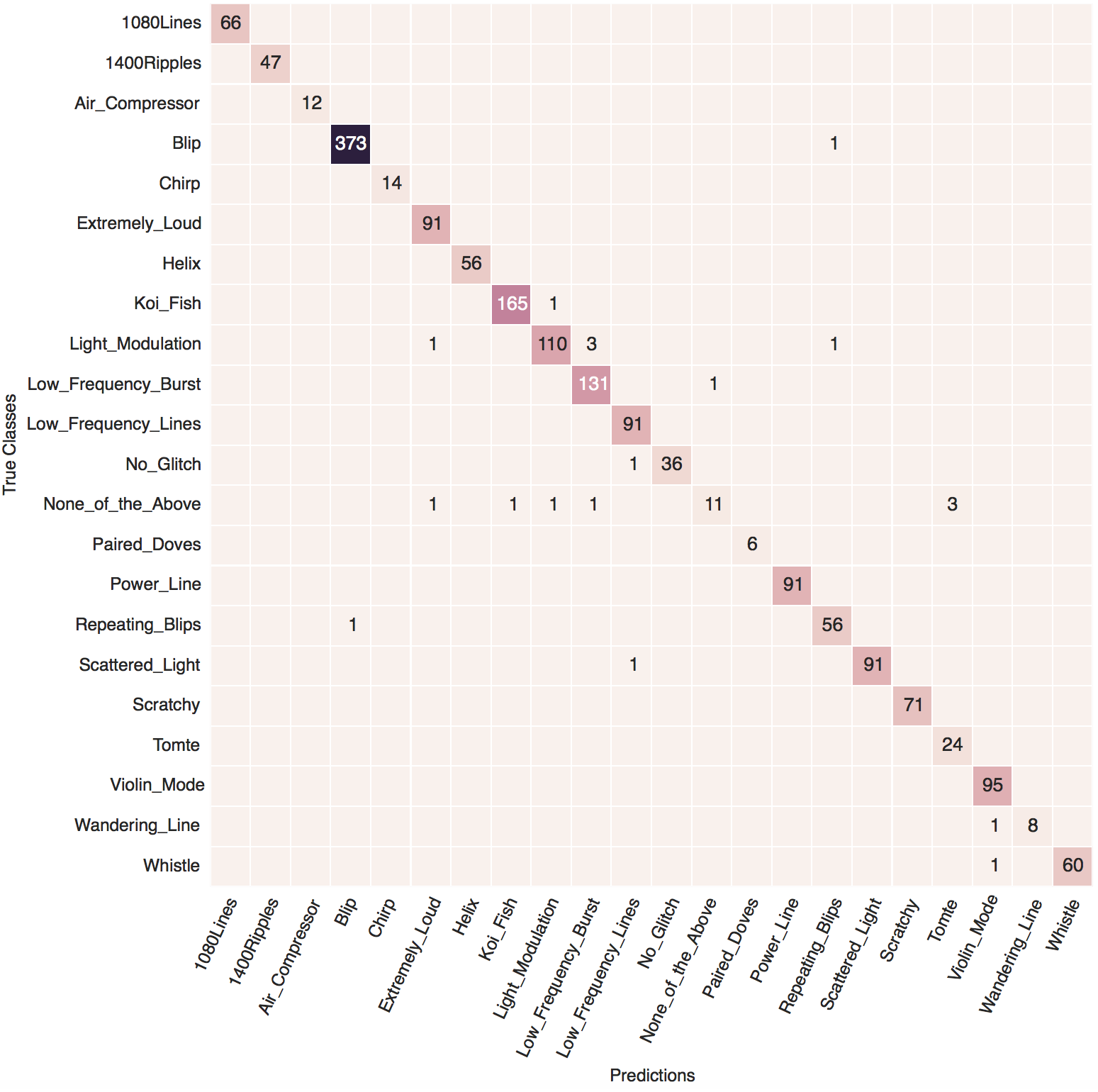}
	\caption{Confusion matrix for InceptionV3. The accuracy is 98.84\% for classification on the test set. The \texttt{Chirp} glitch class (which is also the class that would contain true GW signals from mergers of black hole or neutron star binaries) and the \texttt{Paired\_Doves} class (the smallest class, which only had 24 training examples) was identified with perfect precision and recall, i.e., perfect accuracy.}
	\label{fig:confusionGoogv3}
\end{figure*}

\section{Conclusion}
\label{end}

In this article, we have developed state-of-the-art CNNs for LIGO glitch classification using the \texttt{Gravity Spy} dataset. We have shown that by applying transfer learning from ImageNet, we can obtain excellent results with small training datasets and achieve significantly better the accuracy compared to CNNs trained from scratch only on glitch spectrograms. Furthermore, the training time is significantly reduced with our approach by several orders of magnitude, and the effort required to design CNN models and optimize their hyper-parameters can be eliminated. The algorithms we have introduced in this paper may be used to classify new time-series data in the \texttt{Gravity Spy} project, as well as data streams in real-time from future LIGO and Virgo observing runs as well as KAGRA~\cite{Hiroshe:2014} and LIGO-India~\cite{Unni:2013}, as they come online in the next few years. The transfer learning method also allows us to use the fine-tuned CNNs as feature extractors for clustering algorithms to find new classes of glitches and signals in an unsupervised manner or to label them rapidly in a semi-supervised manner. We expect our methods for glitch classification and clustering may help in finding the instrumental or environmental sources of many classes of glitches whose origins remain unknown, prevent false detection, and enhance the quality of data from gravitational wave detectors thus enabling new scientific discoveries. Furthermore, we anticipate that these techniques may also be useful in general for detecting, classifying, and clustering anomalies in other disciplines.

\section*{Acknowledgements}
\label{ack}
We thank Gabrielle Allen, Ed Seidel, Scott Coughlin, Vicky Kalogera, Aggelos Katsaggelos, Joshua Smith, Kai Staats, Sara Bahaadini, and Michael Zevin for productive interactions. We thank Kai Staats, Laura Nuttall, the \href{https://www.zooniverse.org/projects/zooniverse/gravity-spy/about/team}{Gravity Spy} team, \href{http://gravity.ncsa.illinois.edu/}{NCSA Gravity Group}, and many others for reviewing this article and providing feedback. We are grateful to NVIDIA for supporting this research by donating four P100 GPUs, to Wolfram Research for offering several \href{https://reference.wolfram.com}{Wolfram Language} (\textit{Mathematica}) licenses used for this work, and to Vlad Kindratenko for providing dedicated access to a high-performance machine at the Innovative Systems Lab at NCSA. We acknowledge the Gravity Spy project and the citizen scientists who participated in it for processing and labeling the raw data from LIGO. We also acknowledge the LIGO collaboration for the use of computational resources and for the feedback from the CBC, DetChar, and MLA working groups.

\bibliographystyle{unsrt}
\bibliography{references}

\end{document}